\documentclass[superscriptaddress,twocolumn]{revtex4-2}
\usepackage{amsmath,amsthm, amssymb,amsfonts}
\usepackage{graphicx,textcomp}
\usepackage{enumerate,subeqnarray}
\usepackage{verbatim,natbib}
\usepackage[utf8]{inputenc}
\usepackage[colorlinks=true, citecolor=red, linkcolor=blue, urlcolor=blue ]{hyperref}
\usepackage{siunitx,bm}
\usepackage[title]{appendix}
\usepackage{lineno}
\usepackage{orcidlink}

\newcommand{\oo}[1]{\mathcal O}
\newcommand{\xhat}{\hat{\mathbf x}}
\newcommand{\yhat}{\hat{\mathbf y}}
\newcommand{\zhat}{\hat{\mathbf z}}

\newcommand{\eb}{\bar\epsilon}

\newcommand{\tvec}{\hat{\bm{t}} }
\newcommand{\nvec}{\hat{\bm{n}} }

\newcommand{\xx}[1]{|\mathbf x|}

\DeclareMathAlphabet{\mathsfbi}{OT1}{\sfdefault}{bx}{sl}
\DeclareMathVersion{sfletters}
\SetSymbolFont{letters}{sfletters}{OML}{ntxsfmi}{b}{it}

\makeatletter
\newcommand{\bmsbilow}[1]{%
	\text{\mathversion{sfletters}$\m@th#1$}%
}

\DeclareRobustCommand{\tensor}[1]{%
	\begingroup
	\ifcat\noexpand #1\relax
	\edef\greek@test{\detokenize{#1}}%
	\edef\greek@test{\expandafter\@cdr\greek@test\@nil}%
	\edef\greek@test{\expandafter\@car\greek@test\@nil}%
	\edef\x{\the\lccode\expandafter`\greek@test}%
	\edef\y{\number\expandafter`\greek@test}%
	\ifnum\x=\y\relax
	\bmsbilow{#1}%
	\else
	\mathsfbi{#1}%
	\fi
	\else
	\mathsfbi{#1}%
	\fi
	\endgroup
}
\makeatother

\begin{document}
	
	\title{Hydrodynamic hovering of swimming bacteria above surfaces}
	
	\author{Pyae Hein Htet  \orcidlink{0000-0001-5068-9828}}
	\affiliation{Department of Applied Mathematics and Theoretical Physics, University of Cambridge,	
		Cambridge CB3 0WA, UK}
	\author{Debasish Das \orcidlink{0000-0003-2365-4720}}
	\email{debasish.das@strath.ac.uk}
	\affiliation{Department of Mathematics and Statistics, University of Strathclyde,26 Richmond St, Glasgow G1 1XH, Scotland, UK}
	\author{Eric Lauga  \orcidlink{0000-0002-8916-2545}}
	\email{e.lauga@damtp.cam.ac.uk}
	\affiliation{Department of Applied Mathematics and Theoretical Physics, University of Cambridge,	
		Cambridge CB3 0WA, UK}
	
	\date{\today}

	\begin{abstract}
		Flagellated bacteria are hydrodynamically attracted to rigid walls, yet {past work shows} 
		a `hovering' state where they swim stably at a finite height above surfaces. We use numerics and theory to reveal the physical origin of hovering. Simulations first show that hovering requires an elongated cell body and results from a tilt away from the wall. Theoretical models then identify two essential asymmetries: the response of width-asymmetric cells to active flows created by length-asymmetric cells. A minimal model reconciles near and far-field hydrodynamics, capturing all key features of hovering. 
		
	\end{abstract}
	
	\maketitle
	
	\newpage
	
	Bacteria, highly abundant on Earth, have evolved to thrive in a variety of complex physical environments~\cite{dykhuizen2005}.  Motile bacteria~\cite{wadhwa2022} can  self-propel in fluids using specialized rotary motors~\cite{berg2004}.  Each motor's rotation is transmitted to a short flexible segment -- an elastic hook -- which, in turn, rotates a  helical flagellum, enabling propulsion~\cite{lauga2016}. This locomotion is crucial to the survival of the cells as they search for favorable chemical environments to grow and reproduce~\cite{wadhams2004}. 
	
	Many species of motile bacteria can  grow on surfaces in the form of biofilms~\cite{wood2006,guttenplan2013} 
	and thus a lot of work has focused on understanding the biophysics of swimming cells  near surfaces. To leading order, a swimming bacterium exerts a force-dipole on the surrounding viscous fluid~\cite{drescher2010,lauga2016}, a fundamental physical model which has been used to rationalize a variety of observations, including the collective motion of interacting bacteria~\cite{dombrowski2004,saintillan2008} {and trapping or scattering of individual cells around obstacles} \cite{spagnolie2015,creppy2019}. 
	Near surfaces, a swimming force-dipole is attracted to the nearest wall purely hydrodynamically~\cite{berke2008,spagnolie2012}, 
	and an interplay between near-field hydrodynamics and steric effects produces rich dynamics~\cite{drescher2010}. 
	
	Smooth-swimming bacteria with elongated cell bodies, such as \textit{E.~coli},  change their swimming trajectories from straight to circular  near surfaces~\cite{frymier1995} due to torques exerted on the cells via hydrodynamic interactions with the wall~\cite{lauga2006,lopez2014}. 
	{The stable configuration depends dramatically on the cell body shape;} for example,  spherical bacteria, such as \textit{T.~majus}, reorient perpendicular to the wall and stop swimming~\cite{petroff2015}.  	
	Numerous studies have employed theoretical and computational models to  investigate the dynamics of bacteria near flat boundaries~\cite{phan1987,giacche2010,watari2010,shum2010,hu2015,das2018,das2019}.   
	Building on early experimental work
	~\cite{frymier1995}, the sub-micron length scales between swimming bacteria and nearby surfaces have only been recently resolved experimentally~\cite{li2008,bianchi2017,bianchi2019}.   
	
	Elongated cells swimming along surfaces may attain a stable `hovering' state, self-propelling at a fixed distance from the wall, as predicted in two previous studies~\cite{giacche2010,shum2010}.	
	Subsequently, this hovering state was revealed in experiments that measured the distance between the bacterial cell body and the substrate when the bacterium is initially oriented parallel to it~\cite{bianchi2019}. These striking results contradict the simple dipolar hydrodynamic model -- in which cells are always attracted to the nearest wall -- and yet appear to occur purely due to hydrodynamic interactions. What is the physical mechanism  responsible for hovering?
	
	
	In this paper, we elucidate the  mechanism that allows cells to self-propel at finite distances above surfaces. Using numerical simulations, we demonstrate that hovering results from a slight tilt of the bacterium away from the wall, resulting in a balance between propulsion away from the wall and hydrodynamic wall attraction. 
	To uncover the physical mechanisms responsible for this tilt, we further develop two {increasingly simplified}   theoretical models, which provide a physical explanation for hovering in terms of two essential geometrical asymmetries between the cell body and the flagella: hovering is due to  the response of width-asymmetric  cells  to  active flows created by length-asymmetric cells.  A minimal model finally reconciles near- and far-field hydrodynamics. 

\begin{figure}
		\centering
		\includegraphics[width=0.48\textwidth]{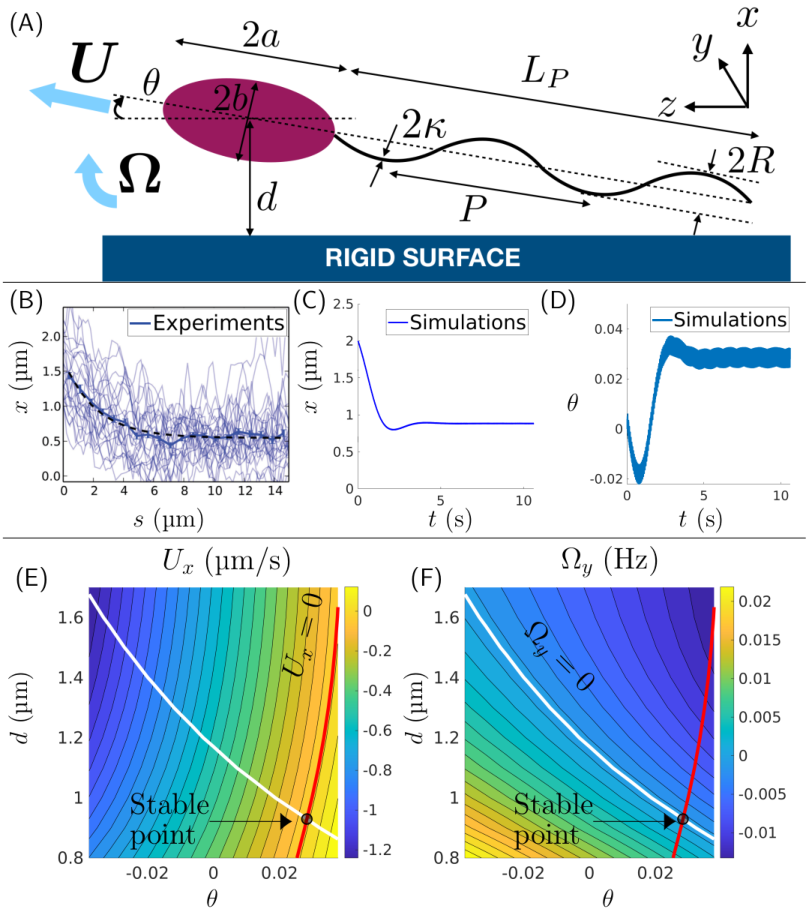}
		\caption{Simulations reveal stable hovering of swimming bacteria with elongated cell bodies. 
			(A): Schematic of a bacterium with an elongated cell body swimming near a rigid surface: cell body of length $2a$ and width $2b$;  helical flagellum of pitch $P$, axial length $L_P$, radius $R$ and cross-sectional radius $\kappa$. The distance between the cell body center and the  surface is $d$ and the flagellar axis is oriented at an angle $\theta$ to the surface. 
			(B): Experimental results for bacteria swimming above a rigid wall, where $s$ is the path length of the trajectory in the $(y,z)$ plane (used with permission of Royal Society of Chemistry, from Ref.~\cite{bianchi2019}).
			(C,D): Distance, $d$, and tilt angle, $\theta$, plotted as a function of time in our dynamic simulations; long-time stable values are $d_\text{eq} =\SI{0.884}{\micro\meter}$ and $\theta_\text{eq} = 0.0286$ rad (see SM~\cite{SM} for movies). 
			(E,F): Contour plots of the wall-normal ($U_x$) and angular velocity ($\Omega_y$) as functions of $d$ and $\theta$, obtained by phase-averaging the fast flagellar rotation. The stable hovering configuration ($d_\text{eq} =\SI{ 0.929}{\micro\meter}$ and $\theta_\text{eq} = 0.0283$) is obtained by the intersection of the nullclines, $U_x=0$ (red line) and $\Omega_y = 0$ (white line).}\label{fig1:schematicnumeric}
	\end{figure}

	The fluid dynamics of bacteria is governed by the Stokes equations which we solve numerically using  boundary elements~\cite{pozrikidis1992,pozrikidis2002,das2018}. The bacterium has a spheroidal cell body of length $2a$ and width $2b$ with dimensions for \textit{E.~coli} as measured  experimentally~\cite{darnton2007} (Fig.~\ref{fig1:schematicnumeric}A and {Supplementary Material} (SM)~\cite{SM}).  The flagella form a left-handed helical bundle that rotates counterclockwise viewed from behind during smooth swimming~\cite{das2019}, modelled as  a single rotating helix aligned with the cell's long axis~\cite{shum2010,das2018,das2019}. The distance between the cell-body center and rigid wall is denoted by $d$ and the tilt angle of the flagellum relative to the wall by $\theta$. 

	We first use dynamic simulations advancing the bacterium's discretized surface numerically in time to reproduce hovering. The stable height above which a bacterium with an elongated cell body swims, measured in a recent experimental work~\cite{bianchi2019} (Fig.~\ref{fig1:schematicnumeric}B), agrees with our simulations (Fig.~\ref{fig1:schematicnumeric}C).  
	Crucially, our  {results} reveal that the swimming bacterium eventually reaches an equilibrium with a small tilt angle $\theta > 0$ away from the wall (Fig.~\ref{fig1:schematicnumeric}D).
	These angles ($\lessapprox$ 1.7 deg), although {likely} too small to be captured experimentally, turn out to play a key role in hovering. 

	We next exploit the instantaneous nature of Stokes flows and the separation of time scales between slow cell dynamics and fast flagellar rotation~\cite{darnton2007} to sweep through the parameter space. For a given cell-to-wall distance ($d$) and tilt angle ($\theta$), we perform several simulations varying the flagellar phase angle from 0 to $2\pi$, and deduce the phase-averaged wall-normal velocity ({$U_x = \dot d$}, contour plot in Fig.~\ref{fig1:schematicnumeric}E) and phase-averaged angular velocity parallel to the surface ({$\Omega_y = \dot \theta$}, see Fig.~\ref{fig1:schematicnumeric}F).

	Hovering corresponds to a dynamic equilibrium i.e.~an intersection of both nullclines in Figs.~\ref{fig1:schematicnumeric}E-F: $U_x=0$ (red line) and  $\Omega_y=0$ (white line). We see   that only one equilibrium point exists, and that it is stable: an increase in $d$ leads to a downward velocity decreasing $d$, and vice versa {(Fig.~\ref{fig1:schematicnumeric}E)}. Similarly, the cell angular velocity  changes sign with $d$ in such a way that  the tilt angle is brought back to its initial direction after any perturbation {(Fig.~\ref{fig1:schematicnumeric}F)}. {There are no other equilibria outside the region illustrated in Figs.~\ref{fig1:schematicnumeric}E-F. If the initial tilt is too high and the cell is pointing away from the wall, it will escape, and if pointing towards the surface, it will crash into it.} Notably, in  contrast to elongated cells, a bacterium with a spherical body cannot hover; instead, it gets attracted all the way to the wall, consistent with previous studies~\cite{das2019} (see SM~\cite{SM}).

	The upward tilt of the cell observed in our simulations suggests that hovering arises from a balance of hydrodynamic attraction resulting  from  force-dipole images~\cite{spagnolie2012} and an apparent repulsion caused by 
	the component of propulsion away from the wall. Hovering is therefore the consequence of tilt. 	To understand its origin, we now develop two increasingly simplified theories. Motivated by the essential role played by  the elongated cell body, we first derive a semi-analytical slender body theory model of the entire bacterium; {this reproduces hovering and allows us to explain the main ingredients required for it to occur.} The model is further reduced to a two-Stokeslet-rod model, revealing the minimal physical ingredients necessary for hovering.
	
\begin{figure}[t!]
		\includegraphics[width=0.48\textwidth]{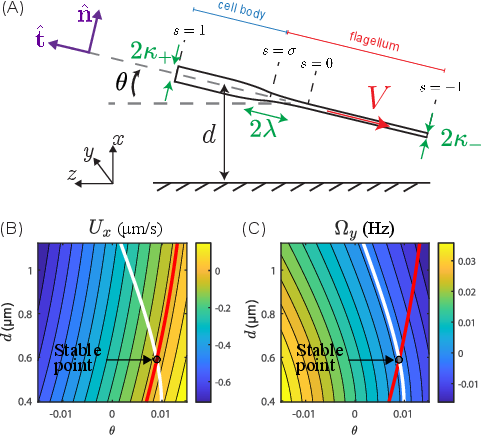}\centering\label{setup}
		\caption{Theoretical model of a slender active bacterium reproduces hovering. (A): Bacterium modelled as an asymmetric active rod of non-uniform radius. (B, C): Contour plots of $U_x$ (B) and $\Omega_y$ (C) plotted against $\theta$ and $d$ {(dimensional quantities)}, as obtained by numerical integration of Eq.~\eqref{UOMEGA};  nullclines $U_x=0$ (red line) and $\Omega_y=0$ (white line). 
		}\label{FIG1}
	\end{figure}

	We thus {first} model the bacterium as a slender cylindrical active rod with an asymmetric shape:  a thick  passive cell body joined to a  thin active flagellum (see sketch in Fig.~\ref{FIG1}A).   
	Denoting by $s$ the arclength along the cell normalised by the half-length $l$ of the whole bacterium, the cross-sectional radius $\kappa(s)$  varies smoothly between $\kappa(s) = \kappa_-$ in the flagellum section $s \in [-1,\sigma - \lambda]$ and $\kappa(s) = \kappa_+$ in the cell body section $s \in [\sigma + \lambda, 1]$, where $2\lambda \ll 1$ is the transition region centered at $s = \sigma$. We define two orientation vectors, $\tvec = \sin\theta \xhat + \cos\theta \zhat$ and $\nvec = \cos\theta \xhat - \sin\theta \zhat$. A slip velocity $\mathbf u_{slip}(s) = -VH_\lambda(\sigma-s)\hat{\mathbf t}$ is prescribed, where $H_\lambda$ is a regularised step function so that $\mathbf u_{slip}$ changes smoothly from $-V$ in $s < \sigma - \lambda$ (directed away from the cell body along the flagellum) to 0 in $s>\sigma + \lambda$ (cell body). The bacterium is inclined at an angle $\theta$ to the wall and its midpoint is at a distance $d$ from the wall. It translates with an instantaneous velocity $\mathbf U = U_x\xhat + U_z\zhat$ and rotates about $s = 0$ with an angular velocity $\Omega_y \yhat$. The surface velocity of the bacterium is given by the kinematic condition, $\mathbf u_{r=\kappa(s)}(s) = \mathbf U + \Omega_y s\nvec - VH_\lambda(\sigma - s)\tvec$.
	
{Henceforth, we scale velocities by $V$, lengths by $l$, and forces by $\mu V l$ (where $\mu$ is the dynamic viscosity) and work in dimensionless variables.}
	The bacterial hydrodynamics is described by slender body theory (SBT)~\cite{hancock1953,batchelor1970,johnson1980,batchelor1970}; this is an integral relation between $\mathbf u(s)$ and the {distribution $\mathbf f(s)$ of point forces (Stokeslets) per unit length} along the bacterium's centerline:
	\begin{equation}\label{SBT}
		{\mathbf u(s) = \tfrac{{1}}{8\pi}\int_{-1}^{1}\mathbf f(s)\cdot[\mathbf G(s; s') + \mathbf G^w(s; s')]\,\mathrm ds'.}
	\end{equation}
	The rigid no-slip surface at $x=0$ is accounted for by hydrodynamic images~\cite{blake1971,blake1974,russel1977}. The tensors $\mathbf G$ and $\mathbf G^w$ represent the flows due to a Stokeslet and its rigid wall image respectively (see SM~\cite{SM}). 
	For a given $V$ and $\kappa(s)$, we determine $\mathbf U$ and $\Omega_y$ using Eq.~\eqref{SBT}, such that the cell always remains force-free ($\int_{-1}^{1}\mathbf f(s)\,\mathrm ds = \boldsymbol 0$) and torque-free ($\int_{-1}^{1}s\tvec\times\mathbf f(s)\,\mathrm ds = \boldsymbol0$).
	
	The dominant contribution to $\mathbf u(s)$ in Eq.~\eqref{SBT} is local and comes from the Stokeslets near $s$ on the centerline~\cite{russel1977}. Evaluation of the integrals using the `divide and conquer' method~\cite{hinch1991} in the limit of small $\kappa$ motivates the choice of an asymptotic variable $\epsilon(s) := \left\{\ln [{2}/{\kappa(s)}]\right\}^{-1}$. Eq.~\eqref{SBT} simplifies to
	\begin{equation}\label{firstapprox}
		{\mathbf u(s) = (4\pi\epsilon(s))^{-1}\left[\tvec\tvec + \tfrac{1}{2}\nvec\nvec\right]\cdot\mathbf f(s) + \mathbf v(s),}
	\end{equation}
	where the non-local contribution $\mathbf v = \mathcal O(f) = \mathcal O(\epsilon)$. Imposing the force- and torque-free, and kinematic conditions, 
	Eq.~\eqref{SBT} yields
	\begin{align}\label{UOMEGA}
		{\begin{pmatrix}
			\eb_0 & 0 & C\eb_1 \\ 0 & \eb_0 & -S\eb_1 \\ C\eb_1 & -S\eb_1 & \eb_2
		\end{pmatrix}\begin{pmatrix}
			U_x \\ U_z \\ \Omega_y
		\end{pmatrix} = \begin{pmatrix}
			S\eb_H +\int_{-1}^{1}\epsilon v_x\,\mathrm ds \\
			C\eb_H + \int_{-1}^{1}\epsilon v_z\,\mathrm ds \\
			\int_{-1}^{1}[C\epsilon v_x- S\epsilon v_z]s\,\mathrm ds
		\end{pmatrix},}
	\end{align}
	where $\eb_H := \int_{-1}^{1}\epsilon(s)H_\lambda(\sigma-s)\,\mathrm ds$ and $\eb_n := \int_{-1}^{1}\epsilon(s)s^n\,\mathrm ds$ {are quantities which depend solely on the bacterial geometry}, $\mathbf v = v_x\xhat + v_z\zhat$, and  {$(C,S) = (\cos \theta, \sin \theta)$}. The solutions to Eq.~\eqref{UOMEGA} are,
	\begin{equation}
		{\mathbf U = {\eb_H}/{\eb_0}\tvec + \mathcal O(\epsilon), \quad \Omega_y = \mathcal O(\epsilon)},
	\end{equation}
	i.e.~the bacterium simply propels forward without rotation with {a speed $\eb_H/\eb_0$} at leading order. We then use Eq.~\eqref{firstapprox} to determine the leading-order force distribution as
	$\mathbf f(s) = 2\pi\epsilon(s)\left[{\eb_H}/{\eb_0} - H_\lambda(\sigma - s)\right]\tvec + \mathcal O(\epsilon^2).$
	At order $\epsilon$, Eqs.~\eqref{SBT}-\eqref{firstapprox}, together with $\mathbf f$, allow us to determine $\mathbf v$ and therefore, $\mathbf U$ and $\Omega_y$.

	The wall-normal translational ($U_x$) and  angular velocity ($\Omega_y$) predicted by this semi-analytical model are shown in Fig.~\ref{FIG1}B-C respectively, with $\lambda=0$. { Here and in what follows we take {$\kappa_- = 0.0028$}; this is the flagellar diameter $0.024$ \textmu{m} divided by the bacterium's length  $2l = 8.68$ \textmu{m} as in the boundary element simulations. Defining $\rho = \kappa_+/\kappa_-$, we focus on  $(\sigma,\rho) = (0.2,30)$, 
	which exhibits hovering. 
	{To plot results as dimensional quantities}, we set the slip velocity to $V = 60$ \textmu{m}/s, which yields a swimming speed of 25 \textmu{m}/s in an unbounded  fluid~\cite{darnton2007}.} 
	Remarkably, this model accurately reproduces the shapes of the nullclines and equilibrium point  in the phase map obtained from numerical simulations in Fig.~\ref{fig1:schematicnumeric}. The SBT approach contains therefore all the necessary physics to explain hovering.

	To reveal the fundamental mechanism of hovering, we next make a further approximation $\theta \ll 1$, as observed in simulations.  Neglecting terms quadratic in $\epsilon$ and $\theta$,  Eq.~\eqref{UOMEGA} yields
	\begin{align}\label{uomegalinear}
		{\begin{pmatrix}
			U_x \\ \Omega_y
		\end{pmatrix} = \begin{pmatrix}
			\tfrac{\eb_H}{\eb_0}\theta \\ 0
		\end{pmatrix}+ \frac{1}{8\pi(\eb_0\eb_2 - \eb_1^2)}\begin{pmatrix}
			\eb_2 & -\eb_1 \\ -\eb_1 & \eb_0
		\end{pmatrix}\begin{pmatrix}
			F_w \\ T_w
		\end{pmatrix},}
	\end{align}	
	where  $F_w \equiv 8\pi\int_{-1}^1\epsilon v_x\,\mathrm ds, T_w  \equiv 8\pi\int_{-1}^1\epsilon v_xs\,\mathrm ds$ are the wall-induced force and torque respectively. 
	We can now derive fully analytical expressions for $v_x$, $F_w$, $T_w$, $U_x$ and $\Omega_y$ (see SM~\cite{SM}).	 
	Eq.~\eqref{uomegalinear} separates the bacterial kinematics into propulsion (first term) and wall effects (second). We illustrate in Fig.~\ref{phyexp}A the wall contribution to $U_x$ and $\Omega_y$ as functions of $d$.
	The wall contribution to $U_x$ is negative for all $d$ (Fig.~\ref{phyexp}A, {blue})  {and balanced by propulsion away from the wall ($\theta>0$) in the hovering state.
		Stability of hovering arises from the correct signs of $\Omega_y$ on either side of the equilibrium height  (Fig.~\ref{phyexp}A, red).		
		\begin{figure}
			\includegraphics[width=0.48\textwidth]{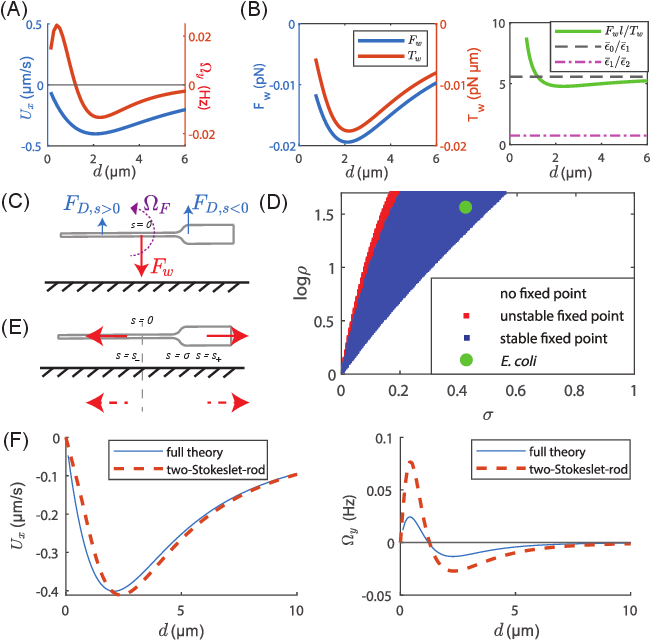}\centering
			\caption{{$U_x, \Omega_y, F_w$ and $T_w$ are plotted against $d$ as dimensional quantities.} (A): Wall contribution to wall-normal translation and angular velocities, $U_x$ (blue) and $\Omega_y$ (red), respectively, plotted against $d$ for $(\sigma,\rho) = (0.2,30)$.  (B): $F_w$ and $T_w$ (left) and their ratio $F_w /T_w$ (right) plotted against $d$. The relative size of $F_w /T_w$ to $\eb_0/\eb_1$ and $\eb_1/\eb_2$ determines the signs of $\Omega_y$ and $U_x$. C): Schematic illustrating the mechanism of hovering: the wall induced force $F_w$ is balanced by unequal drag forces $F_D$ in $s<0$ and $s>0$, resulting in a rotational velocity $\Omega_y$ away from the wall. {(D): Existence and stability of fixed points of the dynamical system $\dot d = U_x, ~\dot \theta = \Omega_y$, for a range of {$(\sigma, \rho)$} values; hovering corresponds to a stable fixed point. The geometrical parameters of {\textit{E.~coli}} lie in the hovering region.} (E): Two asymmetrically placed Stokeslets and their wall images. (F): $U_x$ and $\Omega_y$ plotted against $d$ for $(\sigma,\rho) = (0.2,30)$ as predicted by the full theory (blue solid line) and two-Stokeslets-rod model (red dotted line). 
			}\label{phyexp}
		\end{figure}		
		With either $\sigma = 0$ (flagellum and cell body lengths are equal) or $\rho = 1$ (uniform cross-sectional radius) no hovering is possible. In both these cases,  {$\Omega_y(d) = 0$ has no solution for any $d$, and hence, no equilibrium height exists.} 
		Therefore,  asymmetries in (i) the widths and (ii) the lengths of the cell body/flagellum sections are both essential for hovering. 

		We first investigate the role of width asymmetry, focusing on how $\Omega_y$ and $U_x$ respond to prescribed $F_w$ and $T_w$. 
		The negative $F_w$ and $T_w$ (Fig.~\ref{phyexp}B, left)  experienced by the bacterium  have competing physical effects {on the angular velocity $\Omega_y$}. (I): As the bacterium gets closer to the wall due to an attractive $F_w$ (Fig.~\ref{phyexp}B, left), the section $s>0$ containing the cell body  experiences a larger drag for the same translational velocity than the thinner `flagellum' half $s<0$. This induces a torque rotating the  bacterium  away from the wall (see Fig.~\ref{phyexp}C). (II):  On the other hand,  {the negative wall-induced torque  (Fig.~\ref{phyexp}B, left) rotates the bacterium towards the wall.}  The competition between (I) and (II) is captured by the opposite signs of the two bottom entries of the mobility matrix in Eq.~\eqref{uomegalinear}.  Close to the wall, $F_w$ is sufficiently large {relative to the torque} (i.e.~$F_w/T_w > \eb_0/\eb_1$), so (I) dominates, {and the bacterium rotates away from the wall, whereas the opposite occurs far from the wall (see Fig.~\ref{phyexp}B,  {right; green curve above black curve in near field, and vice-versa away}). (Note we have reintroduced dimensions to $F_w$ and $T_w$ in the plots and therefore labelled the green curve as $F_wl/T_w$.)}
		
		These observations explain why a width asymmetry -- {or more specifically, the asymmetry in hydrodynamic resistances induced by the width asymmetry} -- is essential for hovering. The torque induced by (I) relies on {the difference in hydrodynamic resistance resulting from the width asymmetry}, and thus a  {width-symmetric}  bacterium  only experiences  (II) and   always rotates towards the wall. The translation-rotation coupling in  (I) is captured by the off-diagonal element, $-\eb_1$, of the matrix in Eq.~\eqref{uomegalinear}, which vanishes for a uniform width ($\rho = 1$). 
		
		A similar argument applies for the effects of $F_w$ and $T_w$ on the wall-normal translational velocity, $U_x$.} 
	A negative torque acting on a width-asymmetric cell induces wall repulsion due to a difference in hydrodynamic resistance, but the direct wall attraction induced by $F_w$ is always stronger: $F_w/T_w > \eb_1/\eb_2$ for all $d$ (see top row of Eq.~\eqref{uomegalinear}); this is illustrated in Fig. \ref{phyexp}B (right; green curve always above magenta line).

	{We now investigate the role of length asymmetry (via $\sigma$) on hovering and focus on the image flows that give rise to $F_w$ and $T_w$}. A stability analysis reveals that for given $\rho$, hovering 
	occurs for intermediate values of $\sigma$ (phase diagram in Fig.~\ref{phyexp}D). A length-symmetric bacterium ($\sigma = 0$) fails to hover. {We further illustrate how the hovering height and angle vary with $(\rho, \sigma)$ in SM \cite{SM}.}

	To interpret the role of length asymmetry, we develop a minimal two-Stokeslet-rod model, building on past work~\cite{drescher2011}. The image flows associated with the force distribution along the cell can be qualitatively reproduced by placing one Stokeslet each at  $s_\pm = (\sigma \pm 1)/2$, the  midpoints of the cell body and flagellum respectively (see Fig.~\ref{phyexp}E); note these flows depend only on the length asymmetry. 
	However, two Stokeslets are not sufficient to reproduce hovering; the {hydrodynamic} response of a width-asymmetric cell to these flows is also necessary. 
	We set each of the Stokeslets' strength equal to the magnitude of the force applied by the flagellum (or the cell body) onto the fluid, and describe the hydrodynamics of the asymmetric rod subject to these image flows using resistive force theory (RFT) with resistance coefficients  $\xi_\perp = 2\xi_\parallel=4\pi\mu\epsilon(s)$ ($\epsilon(s)$ depends on the width $\rho$). The predictions from this model qualitatively match all results computed using the full SBT (see Fig.~\ref{phyexp}F for plots of $U_x$ and $\Omega_y$).  {Furthermore, if we subject a length-symmetric bacterium ($\sigma=0$) to the image flows that a length-asymmetric bacterium ($\sigma>0$) would generate, we still observe hovering, provided the width asymmetry is present ($\rho>1$), reinforcing the different roles played by the two asymmetries.
		Our two-Stokeslet-rod model thus fully elucidates the origin of hovering as   the 	{hydrodynamic} response of a width-asymmetric cell to   the flows generated by a length-asymmetric bacterium.}

	In summary, we investigated the hovering of  bacteria swimming above  walls. Numerical simulations revealed that an elongated cell body is required for hovering and that it arises from the apparent wall repulsion due to a slight tilt of the cell away from the wall balancing the well-known hydrodynamic wall attraction of self-propelled cells.  
		{Intriguingly,  our simulations also predict that bacteria with non-slender (spherical) cell bodies do not hover,  the origin of which remains  an open question, since the models  developed here to understand hovering explicitly rely  on the assumption of slender shapes.}

	{Our theoretical model of the bacterium as a slender rod of non-uniform radius, with the near-wall hydrodynamics solved asymptotically using SBT, reproduced all features of hovering from simulations. This model showed that two geometrical asymmetries in the bacterium are essential for hovering:}  the  flows (and their hydrodynamic images) due to a length-asymmetric bacterium, and  the {hydrodynamic}  response of a width-asymmetric bacterium to these flows.  {We further developed a minimal model of asymmetrically placed Stokeslets (length asymmetry) acting on a rod consisting of two sections with different thicknesses (width asymmetry). The two-Stokeslet-rod model reduces to a force-dipole in the far-field, but also reproduces the near-field phenomenon of hovering. 
	
	The phase diagram in Fig.~\ref{phyexp}D predicts that hovering is possible only for an intermediate range of flagellum lengths relative to the cell body. \textit{E.~coli} lies within this range, rationalizing experimental observations~\cite{bianchi2019}.  Given the generic nature of the  mechanism, we expect to see hovering in other bacteria or motile microorganisms that are appropriately {slender and}  asymmetric.

	\end{document}